\documentclass[12pt]{article}
\usepackage{epsfig,graphicx,psfrag}
\usepackage{slashed} 
\newcommand{\be}{\begin{equation}}
\newcommand{\ee}{\end{equation}}
\newcommand{\ba}{\begin{eqnarray}}
\newcommand{\ea}{\end{eqnarray}}

\newcommand{\al}{\alpha}

\newcommand{\GeV}{\mathop{\rm GeV}\nolimits}

\newcommand{\ice}[1]{\relax}

\newcommand{\MSbar}{\overline{\rm MS}}
\newcommand{\MSsch}{{\overline{\rm MS}}}

\newcommand{\pfrac}[2]{\left(\frac{#1}{#2}\right)}

\usepackage{color}
\usepackage{amsmath}

\usepackage{booktabs} %fancy tables
\begin{document}
\thispagestyle{empty}

\phantom{}
\vspace{.2cm}

\begin{flushright}
MZ-TH/12-53\\
SI-HEP-2012-22\\
December 2012
\end{flushright}
\begin{center}
{\Large\bf Understanding perturbative results\\[7pt]
for decays of $\tau$ leptons into hadrons\footnote{Contribution to the 
Festschrift published on the occasion of the 60th birthday of Dmitri Kazakov}}

\vspace{1cm}

{\large \bf S.~Groote$^{1,2}$, J.G.~K\"orner$^2$ and
  A.A.~Pivovarov$^{3,4}$}\\[.4truecm]
$^1$F\"u\"usika Instituut, Tartu \"Ulikool,
  T\"ahe 4, EE-51010 Tartu, Estonia\\[.3truecm]
$^2$PRISMA Cluster of Excellence,\\
  Institut f\"ur Physik der Johannes-Gutenberg-Universit\"at,\\
  Staudinger Weg 7, D-55099 Mainz, Germany\\[.3truecm]
$^3$Institute for Nuclear Research of the\\
  Russian Academy of Sciences, Moscow 117312, Russia\\[.3truecm]
$^4$Department Physik der Universit\"at Siegen,\\
  Walter-Flex-Str.~3, D-57068 Siegen, Germany
\end{center}

\begin{abstract}
We review some results obtained by us and others concerning the 
structure of higher order perturbation series in perturbative QCD and their 
resummation using the renormalization group equation. We illustrate our 
results by a number of examples involving hadronic $\tau$ decays and $e^+e^-$
annihilation.
\end{abstract}

\newpage

\tableofcontents

\newpage
\section{Introduction}
The mathematical framework of Quantum Field Theory (QFT) is a powerful 
tool for the description of elementary particle interactions. The concept of
renormalization and the technique of the $R$-operation generate
renormalizable models of QFT as self-consistent tools for a phenomenological 
analysis~\cite{Bogoliubov:1957gp}. The Standard Model (SM) of ``all
interactions'' is formulated entirely as a local quantum field theoretical
model. The proof of its renormalizability has required some extension of
the standard techniques~\cite{'tHooft:1972fi,Belokurov:1974xe}. Despite the
success in explaining the properties of interactions at high
energies, a closed-form solution of the SM is still missing. The main route
for a quantitative investigation is still the use of perturbation theory (PT) 
where one expands in the small coupling constant where the coefficients of
the perturbation series are computed in terms of integrals represented by
Feynman diagrams.

The classical example of the perturbation theory approach has been developed 
in the context of QED which is part of the SM. After renormalization of QED
there are two parameters -- the electron mass and the fine structure constant
$\alpha$ -- which are to be fitted to observables in terms of a perturbation
series in $\alpha$. With $\alpha=1/137$ being small, the precision of the
results is impressive. However, the question has been raised about the
behaviour of PT series at high orders. Arguments have been put forward that
the series is merely an asymptotic one~\cite{Dyson:1952tj,Parisi:1978bj}. This
has caused much interest in the summation of asymptotic series in
QFT~\cite{Kazakov:1980rd}. Different theoretical aspects of the problem have
been discussed in Ref.~\cite{Kazakov:2002jh}. A quantitative investigation of
the perturbation series in QFT and the question of its convergence has been
presented in Ref.~\cite{Kazakov:1997bt}.

As concerns the problem of PT convergence it has been realized that one can
identify some ``important'' terms in the entire PT series and resum them.
For example, ``large logs'' have first been resummed in~\cite{Landau:1954}. 
These first results even provoked extreme statements about the fate of local
QFT~\cite{Landau:1955ip}. The correct interpretation of the ``resummed''
results emerged within the technique of the renormalization
group~\cite{GellMann:1954fq} (as a review, see Ref.~\cite{Kazakov:2009tr}).
Different types of resummation have been suggested later on: the analysis of
classical solutions~\cite{Lipatov:1976ny}, the famous example of the $1/N_c$
expansion in QCD~\cite{'tHooft:1973jz}, the large $\beta_0$ limit in
QED~\cite{Coquereaux:1980vp}, and the naive nonabelianization in
QCD~\cite{Broadhurst:1994se}.

A precision analysis of the SM requires an extensive use of the
renormalization group (RG). The energy scales of SM are quite different and
the coupling constants (especially $\alpha_s$) are quite large. Indeed, the
electroweak scale given by $M_{Z,W}\sim 90\GeV$ and $m_t=175\GeV$ is quite
high while the hadronic scales are rather low as e.g.\ given by $m_b=5\GeV$
and the light hadron masses at around $1\GeV$. The process of deep inelastic
scattering allows one to probe a wide range of scales by varying $Q$. Modern
experiments include hadrons (LHC) that requires the analysis of hadronization
and the use of QCD at this scale with $\alpha_s(1\GeV)=0.45$. Thus the use of
PT in $\alpha_s(\mu)$ at low energies $\mu=1\GeV$ requires resummation as the
convergence is slow.

Let us list a few technical issues used in the perturbative description of the 
phenomenology of the SM:
\begin{itemize}
\item[i)]Use of higher order PT corrections including NNLO
  corrections which has almost become a standard;
\item[ii)]Precise definition of the expansion parameter or the choice of
  the renormalization scheme. The $\MSbar$-scheme is a standard for technical
  reasons since dimensional regularization has become the main computational
  framework for the calculation of radiative corrections. However, other more 
  physical schemes are also in use such as $\alpha_V$ from the Coloumb part of
  the potential of heavy quarks when using nonrelativistic QCD to describe the
  dynamics of heavy quark production near threshold;
\item[iii)]Resummation of some infinite subsets of the perturbation series in
  $\alpha_s$.
\end{itemize}
The most popular way of improving perturbation theory is the use of the RG 
to sum powers of e.g.\ logs $(\alpha_s\ln(Q/\mu))^n$, 
$\alpha_s(\alpha_s\ln(Q/\mu))^n$. But some other ways are also used: the
$\beta_0$ dominance (naive nonabelianization) which sums terms of the form
$(\beta_0\alpha_s)^n$, or effects of analytic continuation between Euclidean
and Minkowskian regions that basically deal with terms
$(\pi\beta_0\alpha_s)^{2n}$~\cite{Krasnikov:1982fx,Radyushkin:1982kg,%
Pivovarov:1991rh}.

In case of using infinite subsets of the perturbation series, PT allows for
nonpolynomial terms in $\alpha_s$. The PT series are asymptotic and
resummation may provide terms that interfere with non-perturbative expansions
for the description of some processes that account for terms of the form
$\exp{-(1/\alpha_s(Q))}$: higher twists in light-cone type expansions, or
condensate type terms for the expansion of correlation functions at short
distance. This means that numerical values of such non-perturbative parameters
depends on how the PT series are treated.

This is very important for the analysis of hadronic $\tau$ decays. The
theoretical description is simple and related to $e^+e^-$ annihilation. An
unprecedented number of PT terms are available. It is clear that the structure
of the series is important. Also it is not an academic exercise as it is
important for hadronic contributions to $\alpha_{EM}(M_Z)$
\cite{Dolinsky:1991vq,Korner:1998ke} and to the muon $g-2$
\cite{Eidelman:1995ny,Pivovarov:2001mw} which are important for constraints on
the Higgs mass and are also key players in constraining new physics search
beyond the SM~\cite{deBoer:2001nu}. And, it is also a field of interest of
Dima Kazakov.

\section{High order perturbation theory in QCD}
A classic example of application of the PT series in QCD is the analysis of the
$e^+e^-$ cross section~\cite{Bernard:1975cd,Shankar:1977ap,Chetyrkin:1978ta}.
The relevant expansion parameter is large, such that a requisite accuracy
requires a high order PT expansion. With such a large number of terms one may
already encounter the asymptotic nature of the perturbation series in which
case no further increase of precision is possible. The main problem for the
theory is convergence and the interpretation of the numerical values given by
the series. An additional freedom and also complication is that the expansion
parameter is not uniquely determined and the series should be analyzed in a
scheme-invariant way~\cite{Kazakov:1985qh,Gupta:1990jq}. Because of the 
freedom to redefine the scheme it is difficult to judge the quality of
convergence of the series. In this section we present a way to bypass this
complication by establishing a relation between observables.

\subsection{Comparison of observables in 
$e^+e^-$ annihilation and $\tau$ decays}
Within massless perturbative QCD the same Green's function determines the
hadronic contribution to the $\tau$-decay width and the moments of the
$e^+e^-$ cross section. This allows one to obtain relations between physical
observables in the two processes up to an unprecedented high order of
perturbative QCD~\cite{Groote:1997rx}. A precision measurement of the $\tau$ 
decay width allows one then to predict with high accuracy the first few
moments of the spectral density in $e^+e^-$ annihilation integrated up to
$s=m_\tau^2$. 

The question of numerical convergence is influenced to a large extent by the
freedom of choosing the renormalization scheme for the truncated
perturbation series~\cite{Grunberg:1980ja,Stevenson:1981vj,Brodsky:1982gc}.
Therefore, it is desirable to obtain predictions for observables which are
renormalization scheme independent.

We compare moments of the spectral density in $e^+e^-$ annihilation and the
hadronic contributions to $\Gamma(\tau\rightarrow\nu_\tau+{\rm hadrons})$
\cite{Braaten:1988hc,Braaten:1988ea,Narison:1988ni,Braaten:1991qm}. The
reduced decay width $r_\tau$ appearing in
\begin{equation}\label{eqn1}
R_\tau=\frac{\Gamma(\tau\rightarrow\nu_\tau+{\rm hadrons})}%
  {\Gamma(\tau\rightarrow\nu_\tau+\mu+\bar\nu_\mu)}=3(1+r_\tau)
\end{equation}
is determined by massless perturbative QCD, for which the axial and vector
contributions are identical. The expansion for $r_\tau$ starts directly with
$a(\mu^2)$, where $a=\alpha_s/\pi$. In $e^+e^-$ annihilation the cross section
is determined by the imaginary part of the vacuum polarization function,
\begin{equation}
R_{e^+e^-}(s)=12\pi{Im\,}\Pi(s)=N_c\sum Q_i^2(1+r(s))=2(1+r(s)).
\end{equation}
In perturbative QCD one has
\begin{eqnarray}
r(s)&=&a(\mu^2)+(k_1+\beta_0L)a^2(\mu^2)+\Big(k_2-\frac{1}{3}\pi^2\beta_0^2
  +(2\beta_0k_1+\beta_1)L+\beta_0^2L^2\Big)a^3(\mu^2)\nonumber\\&&
  +\Big(k_3-\pi^2\beta_0^2k_1-\frac{5}{6}\pi^2\beta_0\beta_1
  +(3\beta_0k_2+2\beta_1k_1+\beta_2-\pi^2\beta_0^3)L\nonumber\\&&
  +(3\beta_0k_1+\frac{5}{2}\beta_1)L^2+\beta_0^3L^3\Big)a^4(\mu^2)+\ldots
\end{eqnarray}
with $L=\ln(\mu^2/s)$.
We define moments of $r(s)$,
\begin{equation}
r_n(s_0)=(n+1)\int_0^{s_0}\frac{ds}{s_0}\left(\frac{s}{s_0}\right)^nr(s)
\end{equation}
such that
\begin{equation}\label{eqn2}
r_\tau=2r_0(m_\tau^2)-2r_2(m_\tau^2)+r_3(m_\tau^2).
\end{equation}
Eq.~(\ref{eqn2}) can be inverted within perturbation theory. One can then 
express
the perturbative representation of one observable, i.e.\ any given 
$e^+e^-$ moment $r_n(m_\tau^2)$, in powers of $r_\tau$ using the 
perturbative expansion
of the $\tau$-decay observable. The strong coupling constant $\alpha_s$ in any
given scheme serves only as an intermediate agent to obtain relations between
physical observables. The reexpression of one perturbative observable through
another is a perfectly legitimate procedure in perturbation theory and the
result is independent of the choice of the renormalization scheme. One finds
\begin{equation}
r_n(m_\tau^2)=f_{0n}r_\tau+f_{1n}r_\tau^2+f_{2n}r_\tau^3
  +f_{3n}r_\tau^4+f_{4n}r_\tau^5+O(r_\tau^6),\label{eqn3}
\end{equation}
where the coefficients $f_{in}$  are given in the Appendix.

For $r_\tau^{\rm exp}=0.216\pm 0.005$ one can investigate the convergence
properties of the series for the first few moments. One obtains
\begin{eqnarray}
r_0/0.216&=&1-0.284-0.069+0.110+\ldots\\
r_1/0.216&=&1-0.527-0.143+0.177+\ldots\\
r_2/0.216&=&1-0.608-0.115+0.269+\ldots\\
r_3/0.216&=&1-0.648-0.091+0.317+\ldots
\end{eqnarray}
One clearly observes the divergent behaviour of the perturbation series for 
the moments $n=0,1,2,3$. Since the relations between different sets of
observables are scheme independent, there is no freedom in the redefinition of
the expansion parameter. One sees very poor convergence. Therefore, the
question is whether the asymptotic regime has already been reached.

\subsection{Signal of asymptotic behaviour in $\tau$ moments}
How can the asymptotic nature of the series reveal itself? With a
redefinition of the charge one can create any type of the series that hides
the true rate of convergence. Therefore, one should work in a scheme invariant
way. We concentrate on the analysis of the $\tau$-system only.

The spectral density has been calculated with a very high degree of accuracy
within perturbation theory  (see e.g. \cite{Chetyrkin:1996ia,Gorishnii:1990vf,%
Chetyrkin:1996ez,Baikov:2008jh}) and has been confronted with experimental
data to a very high degree of precision~\cite{Braaten:1988hc,Braaten:1988ea,%
Narison:1988ni,Braaten:1991qm}. In Ref.~\cite{Korner:1999kw} arguments have
been given that, within the finite order perturbation theory, the ultimate
theoretical precision has been reached already now. The limit of precision
exists due to the asymptotic nature of the perturbation theory series. The
actual magnitude of this limiting precision depends on the numerical value of
the coupling constant which is the expansion parameter. We do not discuss
power corrections here~\cite{Shifman:1978bx,Gorbunov:2004wy}.

The central quantity of interest in the $\tau$ system is the hadronic spectral
density which can be measured in the finite energy interval
$(0,M_\tau=1.777\GeV)$. The appropriate quantities to be analyzed are the
moments. We define moments of the spectral density by
\begin{equation}\label{intmom}
M_n=(n+1)\int_0^1\rho(s)s^nds\equiv 1+m_n
\end{equation}
($M_\tau$ is chosen to be the unity of mass). The invariant content of the
investigation of the spectrum, i.e.\ independent of any definition of the
charge, would be the simultaneous analysis of all the moments.

In order to get rid of artificial scheme-dependent constants in the
perturbation theory expressions for the moments we define an effective
coupling $a(s)$ directly on the physical cut by the relation
\begin{equation}\label{defofa}
\rho(s)=1+a(s).
\end{equation}
All the constants that may appear due to a particular choice of the
renormalization scheme are absorbed into the definition of the effective
charge (see e.g.\
\cite{Grunberg:1980ja,Krasnikov:1981rp,Kataev:1981aw,Dhar:1983py}). When
defining the effective charge directly by $\rho(s)$ itself we obtain
perturbative corrections to the moments only because of the running of
$\alpha_s$. Without $\alpha_s$ running one would have 
\begin{equation}\label{norunmom}
M_n=1+a(M_\tau)\equiv 1+a\quad\mbox{or}\quad m_n\equiv a
\end{equation}
At any given order of PT the running of the coupling $a(s)$ contains only
logarithms of $s$,
\begin{equation}\label{run}
a(s) = a + \beta_0 L a^2 + (\beta_1 L + \beta_0^2 L^2) a^3
+ (\beta_2 L +\frac{5}{2}\beta_1\beta_0 L^2 + \beta_0^3 L^3) a^4 + \ldots
\end{equation}
where $a=a(M_\tau^2)$, $L=\ln(M_\tau^2/s)$. At fixed order of PT the effects
of running die out for large $n$ moments improving the convergence of the
series
\begin{eqnarray}\label{momff}
m_0&=&a + 2.25 a^2 + 14.13 a^3  + 87.66 a^4  + 654.16 a^5,\nonumber \\
m_1&=&a + 1.125 a^2 + 4.531 a^3  + 6.949 a^4 - 64.77a^5,\nonumber \\
m_2&=&a + 0.75 a^2 + 2.458 a^3  - 1.032 a^4  - 68.98 a^5.
\end{eqnarray}
For large $n$ the moments behave better because the infrared region of
integration is suppressed but in high orders they start to diverge. For large
$n$, the coefficients of the series in Eq.~(\ref{momff}) are saturated by the
lowest power of the logarithm.

To suppress experimental errors from the high energy end of the spectrum,
the modified system of moments  
\begin{equation}\label{intmomkl}
\tilde M_{kl}=\frac{(k+1)!}{k!}\int_0^1\rho(s)(1-s)^kds\equiv 1+\tilde m_k
\end{equation}
can be used. The integral in Eq.~(\ref{intmomkl}) is dominated by
contributions from the low scale region. A disadvantage of choosing such
moments is that the factor $(1-s)^k$ enhances the infrared region strongly and
ruins the perturbation theory convergence. As an example one has
\begin{eqnarray}\label{altmom}
\tilde m_{0}&=&a + 2.25 a^2 + 14.13 a^3 + 87.66 a^4 + 654.2 a^5, \nonumber \\
\tilde m_{1}&=&a + 3.375 a^2 + 23.72 a^3 + 168.4 a^4 + 1373.29 a^5 
\end{eqnarray}
which shows a poor convergence. The reason is the contribution of the
logarithmic terms
\begin{equation}\label{logalt}
(k+1)\int_0^1(1-s)^k\ln(1/s)ds=\sum_{j=1}^{k+1}\frac{1}{j}.
\end{equation}
and
\begin{equation}\label{logalt2}
(k+1)\int_0^1(1-s)^k\ln^2(1/s)ds=\left(\sum_{j=1}^{k+1}\frac{1}{j}\right)^2
  +\sum_{j=1}^{k+1}\frac{1}{j^2} 
\end{equation}
which grow as $\ln(k)$ and $\ln^2(k)$ for large $k$.

The large difference in accuracy between the moments $m_0$ and $m_1$ is a
general feature of the moment observables at the fifth order of perturbation
theory: one cannot get a uniform smallness at this order for several moments
at the same time. For any single moment, one can always redefine the charge
and make the series converge well at any desired rate, but then other moments
become bad in terms of this redefined charge. The invariant statement about
the asymptotic growth is that the system of moments $m_n$ including $m_{n=0}$
cannot be treated perturbatively at the fifth order of perturbation theory.

In order to demonstrate this in a scheme invariant way, we choose the second
moment (which is already well convergent) as a definition of our experimental
charge and find
\begin{eqnarray}\label{invmom}
m_{0}&=&m_2 + 1.5 m_2^2 + 9.417 m_2^3 + 59.28 m_2^4 + 457.54 m_2^5,\nonumber\\
m_{1}&=&m_2 + 0.375 m_2^2 + 1.51 m_2^3 + 2.527 m_2^4 - 17.64 m_2^5,\nonumber\\
m_{2}&=&m_2,\nonumber\\ 
m_{3}&=&m_2 - 0.19 m_2^2 - 0.544 m_2^3 + 0.742 m_2^4 + 16.8 m_2^5,\nonumber\\
m_{4}&=&m_2 - 0.3 m_2^2 - 0.803 m_2^3 + 1.69 m_2^4 + 27.2 m_2^5.
\end{eqnarray}
There is no convergence.

\subsection{$\alpha_s$ from the $\tau$ width in a RG invariant way}
Having in mind that the series expansion has reached the ultimate accuracy, we
try to avoid expansions and analyze the system in a different but still concise
way~\cite{Korner:2000xk}. The observation is that any perturbation theory
observable  generates a scale due to dimensional transmutation and this is
its internal scale. It is natural for a numerical analysis (and this is also 
our suggestion) to determine this scale first and then to transform the result
into a $\MSsch$-scheme using renormalization group invariance.

We use the explicit renormalization scheme invariance of the theory to bring
the result of the perturbation theory calculation into a special scheme first,
followed by a numerical analysis in this particular scheme. Only after that 
we transform the obtained numbers into the $\MSsch$ reference-scheme.

A dimensional scale in QCD emerges as a boundary value parameterizing the
evolution trajectory of the coupling constant. The renormalization group
equation 
\begin{equation}\label{RGE1}
\mu^2\frac{d}{d\mu^2}a(\mu^2) = \beta(a(\mu^2)),\quad a=\frac{\alpha}{\pi} 
\end{equation}
is solved by the integral
\begin{equation}\label{LQCD}
\ln\left(\frac{\mu^2}{\Lambda^2}\right) = \Phi(a(\mu^2)) 
  +\int_0^{a(\mu^2)}\left(\frac1{\beta(\xi)}-\frac1{\beta_2(\xi)}\right)d\xi
\end{equation}
with 
\begin{equation}\label{intb2}
\Phi(a) = \frac1{a\beta_0} + \frac{\beta_1}{\beta_0^2} 
  \ln\left(\frac{a\beta_0^2}{\beta_0 + a\beta_1}\right),\quad
  \beta_2(a)=-a^2(\beta_0 + a\beta_1).
\end{equation}
The $\MSsch$-scheme parameter $\Lambda$ is defined through the expansion
\begin{equation}\label{Lser}
a(Q^2) = \frac1{\beta_0 L}\left(1 - \frac{\beta_1}{\beta_0^2} 
  \frac{\ln(L)}{L^2} \right)+ O\left(\frac1{L^3}\right),\quad
  L=\ln\left(\frac{Q^2}{\Lambda^2}\right). 
\end{equation}
The evolution trajectory of the coupling constant is parametrized by the scale
parameter $\Lambda$ and the coefficients of the $\beta$ function $\beta_i$
with $i>2$ (see e.g. Ref.~\cite{Stevenson:1981vj}). The evolution is invariant
under the renormalization group transformation
\begin{equation}\label{kappa}
a \rightarrow a(1 + \kappa_1 a + \kappa_2 a^2 + \kappa_3 a^3 + \ldots)
\end{equation}
with the simultaneous change
\begin{equation}\label{transform}
\Lambda^2 \rightarrow \Lambda^2 e^{-\kappa_1/\beta_0},
\end{equation}
where $\beta_{0,1}$ is left invariant and
\begin{eqnarray}
\beta_2&\rightarrow&\beta_2-\kappa_1^2\beta_0
  +\kappa_2\beta_0-\kappa_1\beta_1\nonumber\\
\beta_3&\rightarrow&\beta_3+4\kappa_1^3\beta_0
  +2\kappa_3\beta_0+\kappa_1^2\beta_1
  -2\kappa_1(3\kappa_2\beta_0+\beta_2).\nonumber
\end{eqnarray}
This invariance is violated at higher orders of the coupling constant because
one has omitted higher orders for the $\beta$ functions. This is the source for
different numerical outputs of analyses in different schemes.

We introduce an effective charge $a_\tau=\delta_P^{th}$
\cite{Grunberg:1980ja,Dhar:1983py} and extract the parameter $\Lambda_\tau$
which is associated with $a_\tau$ through Eq.~(\ref{LQCD}) with an effective
$\beta$ function
\begin{equation}\label{betatau}
\beta_\tau(a_\tau) = - a_\tau^2
  (2.25 + 4 a_\tau - 12.3 a_\tau^2 + 38.1 a_\tau^3).
\end{equation}
In this procedure the only perturbative objects present are the $\beta$
functions related to $e^+e^-$ annihilation (in Eq.~(\ref{RGE1})) and
$\tau$ decay (see Eq.~(\ref{betatau})). We treat them both as concise
expressions, and at every order of the analysis we use the whole information
of the perturbation theory calculation. For the coupling constant in the
$\MSsch$ scheme we finally find 
\begin{equation}\label{theoryNNLO}
\al_s=0.3184 \pm 0.0159.
\end{equation}
For the reference value of the coupling constant at the scale
$M_Z = 91.187\GeV$ we run to this reference scale with the four-loop $\beta$
function in the $\MSsch$ scheme~\cite{vanRitbergen:1997va} and three-loop
matching conditions at the heavy quark (charm and bottom)
thresholds~\cite{Chetyrkin:1997sg} to get~\cite{Korner:2000xk}
\begin{equation}\label{rgi0}
\al_{s}(M_Z) = 0.1184 \pm 0.0007_{exp} \pm 0.0006_{cb} 
\end{equation}
Can one do better than finite order PT? Yes, but then one has to resum!

\section{Resummation on the cut: $q^2>0$}
Since new higher order terms in the perturbation series will not become
available in the near future for $e^+e^-$ annihilation or the 
$\tau$ decays, it is
tempting to speculate on the general structure of the series within
PT~\cite{'tHooft:1977am,David:1982qv,Mueller:1984vh}. Much attention has been
recently paid to possible factorial divergences in the PT series
\cite{Zakharov:1992bx,Beneke:1992ch,Ball:1995ni,Bigi:1994em,Manohar:1994kq}
which is generated through the integration over an infrared region in momentum
space~\cite{Webber:1994cp}.

Going beyond finite order PT directly by taking into account RG logs in 
$\rho(s)$ does not work. Using a simple approximation $\rho(t)=\alpha_s(t)$ 
and 
\[
\alpha_s(t)={\frac{\alpha_s(s)}{1-\beta_0\alpha_s(s)\ln(s/t)}}
=\alpha_s(s)\sum_{n=0}^\infty (\beta_0\alpha_s(s)\ln(s/t))^n
\]
one finds
\[
 F(s)=\frac{1}{s}\int_0^s \alpha_s(t)dt=\alpha_s(s)
\sum_{n=0}^\infty (\beta_0\alpha_s(s))^n n!
\]
with factorial growth which is not Borel summable. The reason is the Landau
pole in the expression for $\alpha_s(t)$ or the divergence of the integrand
outside the convergence circle $|a(s)\ln(s/t)|<1$. Higher order terms in
$\rho(t)$ are important~\cite{Krasnikov:1995is}.

For an estimate of the uncertainties of the theoretical predictions for the
$\tau$ lepton width, different approaches have been used for the definition 
of the integration over the infrared region. This problem has been widely 
discussed in the
literature (see, e.g.\ Ref.~\cite{Altarelli:1994vz}). We propose a set of
schemes that regularize the infrared behaviour of the coupling constant in
general and allow one to use any reference scheme for the high energy
domain~\cite{Krasnikov:1995is}. All these schemes are perturbatively
equivalent at high energies. The uncertainties that come from the low energy
region are quite essential as our study will show.

\subsection{Example with an explicit solution for $\alpha_s(\mu^2)$}
Consider first an example with an explicit solution for
$\alpha_s(\mu^2)$~\cite{Krasnikov:1995is}. Consider a $\beta$ function
\[
\beta(a)=-\frac{a^2}{1+\kappa a^2}, \quad \kappa>0
\]
with $\beta^{as}(a)=-a^2+\ldots$ at $a\to 0$. The RG equation has a solution
\[
a(\mu^2)=\frac1{2\kappa}\left(-\ln\pfrac{\mu^2}{\Lambda^2}
+\sqrt{\ln^2\pfrac{\mu^2}{\Lambda^2}+4\kappa}\right),
\]
and the pole at $\mu^2=\Lambda^2$ of the asymptotic solution
$a^{as}(\mu^2)=(\ln(\mu^2/ \Lambda^2))^{-1}$ disappears.

The example shows that a particular way of summing an infinite number of 
specific
perturbative terms for the $\beta$ function can cure the Landau pole problem.
No nonperturbative terms are added but the freedom of choosing a
renormalization scheme for an infinite series was used. This result can be 
considered
either as a pure PT result in some particular RGE after an infinite resummation
or as a sort of Pad\'e approximation of some real $\beta$ function that may
also include nonperturbative terms. The only important point is
that the running coupling obeying the RG equation with such a $\beta$ function
has a smooth continuation to the infrared region. Because the expansion
parameter becomes large in the infrared region, the polynomial approximation
is invalid in this domain. This is a particular example of the general
situation that the expansion in the unphysical parameter $\alpha_s$ is 
incorrect and the appropriate way to proceed is to expand one physical 
quantity through another.

\subsection{Example with an explicit expression\\
  for the integral over the effective charge}
Consider an example with
\[
\beta(a) = \frac{-a^2}{1+2a}.
\]
In this case the integral
\[
F(s)=\frac{1}{s}\int_0^s a(t)dt
\]
can be found explicitly~\cite{Krasnikov:1995is}. The RG equation for the
effective charge $a(s)$ is given by
\[
\ln(s/\Lambda^2)=\frac{1}{a(s)}-2\ln a(s),
\]
and $F(s)$ reads
\[
F(s)=\frac{1}{s}\int_0^s a(t)dt=  a(s)+a(s)^2-a(s)^2\exp(-\frac{1}{a(s)}).
\]
The last term gives the ``condensate'' contribution up to logarithmic
corrections $F^{cond}(s)\sim \Lambda^2/s$. This example shows that a change of
the evolution in the infrared region resums factorials. The last term cannot
be detected by integrating the series near $a=0$.

In order to study higher orders of PT, the Borel transformation is often used.
The Borel analysis for this example reads
\begin{equation}\label{bordef}
F(a)=\int_0^\infty e^{-\xi/a}B(\xi)d\xi
\end{equation}
and the Borel image $B(x)$ is $B(x)=1 + x + (1-x)\theta(x-1)$. The PT series
with all coefficients known does not allow one to restore the exact answer
through Borel summation. A naive Borel image for polynomials from the 
definition 
(\ref{bordef}) behaves as $B(x) = \sum_{k=1}f_k x^k/(k-1)!$ which is correct
at small $x$ but not at large $x$. The explicit result shows that the Borel
image is singular at $x=1$.

There is still another possibility to treat the PT series: resummation in the
complex $q^2$ plane~\cite{Pivovarov:1991rh}.

\section{Resummation on the contour}
By integrating the function $\Pi^{\rm had}(z)$ over a contour in the complex
$q^2$ plane above the physical cut $s>0$ one finds
\[
\oint_C \Pi(z)dz = \int_{\rm cut} \rho(s) ds
\]
with 
\[
\rho(s)=\frac{1}{2\pi i}\left( \Pi(s+i0)- \Pi(s-i0)\right).
\]
One then uses the approximation
$\Pi^{\rm had}(z)|_{z\in C} \approx \Pi^{\rm PT}(z)|_{z\in C}$
which is well justified when one is sufficiently far from the physical cut. 
One then obtains
\[
\oint_C \Pi^{\rm had}(z)dz = \int_{\rm cut} \rho(s) ds
= \oint_C \Pi^{\rm PT}(z)dz
\]
where the integral over the hadronic spectrum is computable in perturbative 
QCD.
The total decay rate of the $\tau$ and its moments are the quantities that can
be computed in this way. The use of the RG improved $\Pi^{\rm PT}(z)$ on the 
contour
has been first considered in~\cite{Pivovarov:1991rh}. The technique is now
known as Contour Improved Perturbation Theory 
(CIPT)~\cite{Pivovarov:1991rh,Le Diberder:1992te}. Parametrizing the
contour by $Q^2=M_\tau^2e^{i\varphi}$ one obtains for $M_{kl}$
\[
M_{kl} = 1+m_{kl} = \frac{(-1)^l}{2\pi}\frac{(k+l+1)!}{k!l!}\int_{-\pi}^\pi
  \Pi(M_\tau^2e^{i\varphi})(1+e^{i\varphi})^ke^{i(l+1)\varphi}d\varphi.
\]
This program has been realized for $\tau$ decays and the extraction of 
$\alpha_s(M_\tau^2)$ and $m_s(M_\tau^2)$ in a series of 
papers~\cite{Groote:1997cn,Groote:1997kh,Korner:2000wd}.

\subsection{Resummation using the four-loop $\beta$ function}
The resummation of the PT series for observables related to $\tau$ decays with
the four-loop $\MSbar$ $\beta$ function has been studied in
Ref.~\cite{Groote:1997cn}. The integrals over the contour give the resummed
functions that are analytic at the origin with a finite radius of convergence.
The convergence radius of a series given by an integration over
the contour and can be analyzed through the singularity structure in the 
complex
$a_\tau$ plane~\cite{Groote:1997cn}. In the lowest order example, the radius
of convergence is determined by the solution of the equation
$1 - i\pi\beta_0 a=0$, leading to a region of convergence given by 
$|a| < 1/\pi\beta_0$. In higher orders of the $\beta$ function the 
convergence properties of the resummed functions
$M_{i,n}(a_\tau,\beta)$ become quite involved. The evolution of $a$ along the
contour $Q^2=m_\tau^2e^{i\phi}$, $\phi\in[-\pi,\pi]$ is governed by the
renormalization group equation
\begin{equation}\label{RGE}
-i\frac{\partial a}{\partial\phi}=\beta(a)=-a^2(1+ca+c_2a^2+c_3a^3+\ldots\ ),
\end{equation}
and the closest singularity in the complex $a_\tau$ plane then determines the 
convergence radius of the resummed functions $M_{i,n}(a_\tau,\beta)$. The
results for critical values of $\alpha_s=\alpha_s(m_\tau^2)$ for increasing 
orders of the $\beta$ function read
\begin{equation}\label{radii}
\alpha_s^{(1)}=0.444,\quad
\alpha_s^{(2)}=0.331,\quad
\alpha_s^{(3)}=0.310,\quad
\alpha_s^{(4)}=0.299.
\end{equation}
The convergence radii become smaller as the order of the $\beta$ function
increases. It is interesting to speculate about the possibility that the 
convergence radius continues to shrink as one goes to ever higher orders of the
$\beta$ function, including the possibility that the convergence
radius shrinks to zero when the order of the perturbative $\beta$ function 
expansion goes to infinity.

The value of $a_\tau$ is outside the convergence region. This means that the
perturbative approximation for the moments diverges at the scale determined by
the experimental data for the semileptonic $\tau$-decay width. The resummed
values are not accessible by using higher and higher order approximations of
PT in polynomial form.

\subsection{Relations between observables using resummation}
Here we discuss a model of how the technique of using the direct PT relations
between observables can give results that are also obtained in a more
sophisticated resummation approach~\cite{Groote:1997kh}. Consider two
observables given by the perturbative series in some given scheme, i.e.
\begin{equation}\label{example1}
f(a)=a(1-a+a^2-\ldots\ )=\frac{a}{1+a}
\end{equation}
and
\begin{equation}\label{example2}
g(a)=a(1-2a+4a^2-\ldots\ )=\frac{a}{1+2a}.
\end{equation}
The functions $f(a)$ and $g(a)$ can be seen to be related by 
\begin{equation}\label{convergent}
g(f)=\frac{f}{1+f}=f(1-f+f^2-\ldots\ ).
\end{equation}
If we fit the right-hand side of Eq.~(\ref{example1}) to an experimental value
of about $f=0.6$, we get $a=1.5$. But for this value of the coupling, the
series in Eq.~(\ref{example1}) diverges. One cannot get $a$ from 
the series without an appropriate resummation procedure which, in this case,
is trivially given by the exact formula. As a consequence we cannot get a
prediction for $g$ using the series in Eq.~(\ref{example2}) in terms of $a$.
On the other hand, the direct relation in terms of the series in
Eq.~(\ref{convergent}) converges perfectly and gives an unambiguous result for
$g$ in terms of the measured $f$. Of course, for such an improvement to occur
one has to analyze in detail the underlying theory and the origin of the
series. The analysis of the $\tau$ system taking into account $m_s$
corrections has been performed in finite order PT in
Refs.~\cite{Maltman:1999rh,Chetyrkin:1998ej}, and with resummation in
Ref.~\cite{Pich:1998yn}. Resummation on the contour along the above lines in
an effective scheme has been done in Ref.~\cite{Korner:2000wd}. It happens
that numerical results of the different techniques are different. It is
important to understand the difference, or to answer the question to what
extent the resummation recipe restores the same correlation function.

\section{Comparison of resummation techniques}
Because of the arbitrariness of resummation, it is important to understand 
the relation between different resummation techniques. Clearly, different 
resummations
of asymptotic series give functions that have the same asymptotic expansion
but differ from the original functions. General lore is that the difference 
between different resummation schemes behaves
as $\exp{(-1/\alpha_s)}$. This form emerges from the Borel resummation recipe
and is confirmed by an explicit example with resummation on the
cut~\cite{Krasnikov:1995is}, although other forms are also
possible~\cite{Penin:1996zk}.

Here we discuss the relation between CIPT and the resummation on the cut known
as analytic PT~\cite{Shirkov:1997wi,Bakulev:2010gm}, following the lines of
Ref.~\cite{Groote:2001im}. At leading order the moments can be expanded in a
convergent series in $\alpha_\tau$ for $\beta_0\alpha_\tau<1$. Within the 
contour technique of resummation the finite
radius of convergence is a general
feature which persists in higher orders of the $\beta$
function~\cite{Groote:1997kh,Groote:1997cn}. The convergence radius decreases
when higher orders of the $\beta$ function are included. In practice, for
$\alpha_\tau^{\rm exp}/\pi=0.14$ the relation
$\alpha_\tau^{\rm exp}<1/\beta_0$ is still approximately valid. The
exact expression provides an analytic continuation beyond the convergence
radius even when $\alpha_\tau$ lies outside the convergence radius. 

We consider the moment $m_{00}$ and proceed with the analysis by constructing
an efficient computational scheme. By integrating $n$ times by parts, one
obtains
\begin{equation}\label{momexpand1}
m_{00}=\frac1{\pi\beta_0}\Bigg\{\phi+\sum_{j=1}^{n-1}\Gamma(j)
  \pfrac{\beta_0\alpha_\tau}{\pi r}^j\sin(j\phi)
  +\frac{\Gamma(n)}2\pfrac{\beta_0\alpha_\tau}\pi^n\int_{-\pi}^\pi
  \frac{e^{i\varphi}d\varphi}{(1+i\beta_0\alpha_\tau\varphi/\pi)^{n}}\Bigg\}
\end{equation}
where the polar coordinate functions $r$ and $\phi$ are defined by
\begin{equation}
1\pm i\beta_0\alpha_\tau=re^{\pm i\phi},\qquad
r=\sqrt{1+\beta_0^2\alpha_\tau^2},\qquad
\phi=\arctan(\beta_0\alpha_\tau).
\end{equation}
The $n$-fold integration by parts has removed a polynomial of order $n$ from 
the expansion of the logarithm.

One gets an asymptotic expansion where the residual term, i.e.\ the last term
in Eq.~(\ref{momexpand1}), is of the formal order $\alpha_\tau^n$. However,
the result is not a series expansion in the original coupling determined in
the Euclidean domain but an expansion over a more complicated system of
functions emerging in analytic PT~\cite{Bakulev:2010gm}. The system of
functions is ordered and the asymptotic expansion is valid in the sense of
Poincar\'e. The system of functions is obtained by using the expression for
the running coupling in the Euclidean domain and continuing it into the
complex plane and onto the cut. When the analytic structure of the initial
function is known, asymptotic expansions which converge fast for the first few
terms (as a representation in the form of Eq.~(\ref{momexpand1})) are useful
for practical calculations. The expansion in Eq.~(\ref{momexpand1}) can give a
better accuracy (for some $n$ and $\alpha_\tau$) than a direct expansion in
$\alpha_s$. Indeed, this expansion includes a partial resummation of the
$\pi^2$ terms which is a consequence of the analytic
continuation~\cite{Krasnikov:1982fx}. Therefore, the expansion can be
understood as being done in terms of quantities defined on the cut. Because
the region near the real axis is important, the continuation causes a change
of the effective expansion parameter
$\alpha_\tau\to\alpha_\tau/\sqrt{1+\beta_0^2\alpha_\tau^2}$. The first term in
the expansion shown in Eq.~(\ref{momexpand1}) is just the value for the
spectral density expressed through the coupling in the Euclidean domain.

With a concise expression for the moments at hand one can change the form of
the residual term. The relation
\begin{eqnarray}\label{end}
\lefteqn{(n-1)!\pfrac{\beta_0\alpha_\tau}\pi^n\int_{-\pi}^\pi
  \frac{e^{i\varphi}d\varphi}{(1+i\beta_0\alpha_\tau\varphi/\pi)^n}
  \ =}\nonumber\\
  &=&2\pi e^{-\pi/\beta_0\alpha_\tau}-(n-1)!\pfrac{\beta_0\alpha_\tau}\pi^n
  \left(\int_{-\infty}^{-\pi}+\int_{\pi}^{\infty}\right)
  \frac{e^{i\varphi}d\varphi}{(1+i\beta_0\alpha_\tau\varphi/\pi)^n}
\end{eqnarray}
is valid for any $n$ and leads to a representation of the zeroth order moment
of the form
\begin{eqnarray}\label{momexpand2}
m_{00}&=&\frac1{\pi\beta_0}\Bigg\{\pi e^{-\pi/\beta_0\alpha_\tau}+\phi
  +\sum_{j=1}^{n-1}(j-1)!\pfrac{\beta_0\alpha_\tau}{\pi r}^j
  \sin(j\phi)\nonumber\\&&\qquad
  -\frac{(n-1)!}2\pfrac{\beta_0\alpha_\tau}\pi^n
  \left(\int_{-\infty}^{-\pi}+\int_\pi^\infty\right)
  \frac{e^{i\varphi}d\varphi}{(1+i\beta_0\alpha_\tau\varphi/\pi)^{n}}
  \Bigg\}.\qquad
\end{eqnarray}
Note that a ``nonperturbative'' term $e^{-\pi/\beta_0a_\tau}$ has appeared.

The moments are analytic functions of $\alpha_\tau$ for small values of the
coupling $\alpha_\tau$. This means that the non-analytic piece in
Eq.~(\ref{end}) cancels the corresponding piece in the residual term. If the
residual term is dropped, the analytic structure drastically changes depending
on which representation, either Eq.~(\ref{momexpand1}) or~(\ref{momexpand2}),
is used. 

\begin{figure}
\begin{center}\epsfig{figure=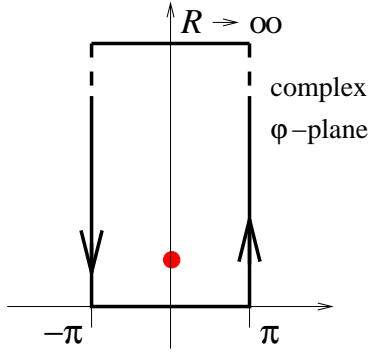,scale=0.7}
\caption{\label{fig7}Integration contour in the complex $\varphi$ plane}
\end{center}
\end{figure}

One can recover the integral form of the moments as integrals over a 
spectral density
by going to the complex plane in $\varphi$ (see Fig.~\ref{fig7})
\begin{equation}
m_{00}=\frac{\alpha_\tau}{2\pi^2}\int_{-\pi}^\pi
  \frac{(1+e^{i\varphi})d\varphi}{1+i\beta_0\alpha_\tau\varphi/\pi}
\end{equation}
This representation is different from Eq.~(\ref{momexpand1}) for $n=1$. The
difference is an integral which can be explicitly computed,
\begin{equation}
\frac{\alpha_\tau}{2\pi^2}\int_{-\pi}^\pi
  \frac{d\varphi}{1+i\beta_0\alpha_\tau\varphi/\pi}
  =\frac{1}{\pi\beta_0}\arctan(\beta_0\alpha_\tau).
\end{equation}

Now we consider the integration over a rectangular contour in the complex
$\varphi$ plane. The part of the contour on the real axis from $-\pi$ to $\pi$
leads to the moment expressions. The integral over the contour is given by 
the residue at
the pole $\varphi=i\pi/\beta_0\alpha_\tau$. We thus have
\begin{equation}
m_{00}=\frac1{\beta_0}(1+e^{-\pi/\beta_0\alpha_\tau})
  -\frac1{\beta_0}\int_0^\infty\frac{(1-e^{-\xi})d\xi}{\pi^2
  +(\xi-\pi/\beta_0\alpha_\tau)^2}.
\end{equation}
With the substitutions $-\pi/\beta_0\alpha_\tau=\ln(\Lambda^2/M_\tau^2)$,
$-\xi=\ln(s/M_\tau^2)$ one obtains
\begin{equation}
m_{00}=\frac1{\beta_0}\left(1+\frac{\Lambda^2}{M_\tau^2}\right)
  -\frac1{\beta_0}\int_0^{M_\tau^2}\frac{(1-s/M_\tau^2)ds}{(\pi^2
  +\ln^2(s/\Lambda^2))s}.
\end{equation}
Finally
\begin{equation}\label{m00further}
m_{00}=\frac1{\beta_0}\pfrac{\Lambda^2}{M_\tau^2}
  +\frac1{\pi\beta_0}\int_0^{M_\tau^2}\arccos
  \left(\frac{\ln(s/\Lambda^2)}{\sqrt{\pi^2+\ln^2(s/\Lambda^2})}\right)
  \frac{ds}{M_\tau^2}.
\end{equation}
One recognizes this representation as an integration over the singularities of
$\Pi(q^2)$. In addition to a cut along the positive semi-axis, there appears
also a part of the singularity on the negative real $s$ axis. This part is a
pure mathematical feature of the definite approximation chosen for $\Pi(q^2)$.
The result reads
\begin{equation}\label{m00result}
m_{00}=\int_{-\Lambda^2}^{M_\tau^2}\frac{\sigma(s)ds}{M_\tau^2}
\end{equation}
with
\begin{equation}\label{sigmadef}
\sigma(s)=\frac1{\beta_0}\theta(\Lambda^2+s)\theta(-s)+\frac1{\pi\beta_0}
  \theta(s)\arccos\left(\frac{\ln(s/\Lambda^2)}{\sqrt{\pi^2
  +\ln^2(s/\Lambda^2})}\right).
\end{equation}
This formal result can be rewritten as an integration over the spectrum
$\sigma(s)$ using Cauchy's theorem. Indeed,
\begin{equation}
{\rm Disc\,}\Pi(s)=\frac{2\pi i}{\beta_0}\Bigg\{\theta(\Lambda^2+s)
  \theta(-s)+\frac1\pi\theta(s)\arccos
  \left(\frac{\ln(s/\Lambda^2)}{\sqrt{\pi^2+\ln^2(s/\Lambda^2})}\right)\Bigg\}
\end{equation} 
which coincides with $\sigma(s)$ in Eq.~(\ref{sigmadef}). The part of the
spectrum on the positive real axis is an analytic continuation of the function
$\Pi(Q^2)$ to the
cut~\cite{Krasnikov:1982fx,Pivovarov:1991rh,Radyushkin:1982kg,Shirkov:1997wi}.
It can be written in the form
\begin{equation}\label{sigc}
\sigma_c(s)=\frac1{\pi\beta_0}\arctan(\beta_0\alpha(s)).
\end{equation}
The differential equation determining the continuum part $\sigma_c(s)$
through its initial value $\sigma(M_\tau^2)$ can be constructed by
differentiating Eq.~(\ref{sigc}) with respect to $s$,
\begin{equation}
s\frac{d\sigma_c(s)}{ds}
  =-\beta_0\pfrac{\alpha(s)}\pi^2\frac1{1+\beta_0^2\alpha(s)^2}
\end{equation}
By inverting Eq.~(\ref{sigc}) one has
$\beta_0\alpha(s)=\tan(\pi\beta_0\sigma_c(s))$ for $s>0$. Therefore, one 
obtains
\begin{equation}\label{sigmacev}
s\frac{d}{ds}\sigma_c(s)=-\frac1{\pi^2\beta_0}\sin^2(\pi\beta_0\sigma_c(s))
  \quad\mbox{for\ }s>0.
\end{equation}
This equation can indeed be considered as an evolution equation for the
spectral density $\sigma_c(s)$ determining $\sigma_c(s)$ through its initial
value $\sigma(M_\tau^2)$. Therefore, one can introduce an effective charge
$\alpha_M(s)=\pi\sigma_c(s)$ with a corresponding evolution equation 
\begin{equation}\label{almev}
s\frac{da_M(s)}{ds}=-\frac1{\pi^2\beta_0}\sin^2(\pi\beta_0a_M(s)).
\end{equation}
Thus, one defines the coupling as the value of the spectral density on the cut
far from the infrared region. The evolution of this coupling is
calculated by taking into account analytic continuation. In this case, the
coupling
has an infrared fixed point with the value $a_M(0)=1/\beta_0$. If Adler's
function starts with another power of the coupling constant, as is the case
for gluonic observables, this picture will change. For 
\begin{equation}
D(Q^2)=\pfrac{\alpha_E(Q^2)}\pi^2
\end{equation}
the spectral density in the leading-order $\beta$-function approximation reads
\begin{equation}
\rho(s)=\frac1{\beta_0^2}\ \frac1{\ln^2(s/\Lambda^2)+\pi^2}
  =\frac{\alpha^2(s)}{\pi^2(1+\beta_0^2\alpha^2(s))}.
\end{equation}
and an effective coupling is
\begin{equation}\label{effM2}
\bar a_M(s)=\frac{\alpha(s)}{\pi\sqrt{1+\beta_0^2\alpha^2(s)}}.
\end{equation}
The $\beta$ function for the effective coupling obtained from
Eq.~(\ref{effM2}),
\begin{equation}
\bar\beta_M(\bar a_M)=-\beta_0\bar a_M^2\sqrt{1-(\pi\beta_0\bar a_M)^2}
  =-\beta_0\bar a_M^2\left(1-\frac12(\pi\beta_0\bar a_M)^2+O(\bar a_M^4)\right)
\end{equation}
differs from the $\beta$ function of the effective coupling $a_M$ (c.f.\
Eq.~(\ref{almev})),
\begin{equation}
\beta_M(a_M)=-\frac1{\pi^2\beta_0}\sin^2\left(\pi\beta_0a_M\right)
  =-\beta_0a_M^2\left(1-\frac{1}{3}(\pi\beta_0a_M)^2+O(a_M^4)\right)
\end{equation}
at next-to-leading order. Thus, resummation on the contour and on the
positive semi-axis of the $s$ plane differs by the integral over the negative
real semi-axis for $s$.

\begin{figure}
\begin{center}
\epsfig{figure=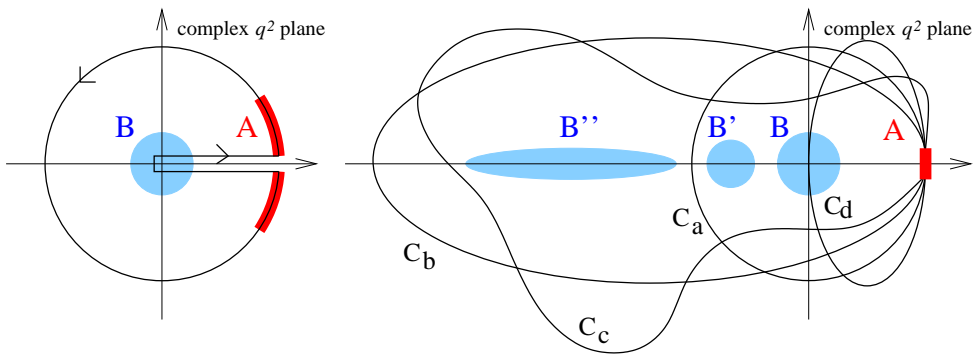, scale=0.7}
\caption{\label{figa}Contours in the complex plane starting and ending in the
  region A. The different contours take into account possible occurrences of 
  singularities. The
  left hand side of the figure shows the standard circular path which
  circumvents the singular region B. Other singular
  regions as discussed in the text (regions B$'$ and B$''$) lead to
  different possibilities for choosing a path (C$_{\rm a}$, C$_{\rm b}$,
  C$_{\rm c}$). The path C$_{\rm d}$ crosses the singular region and,
  therefore, cannot be used from the perturbation theory point of view.}
\end{center}
\end{figure}

In the context of the contour formulation it is not essential what particular 
point-by-point
behaviour exists in the infrared region. For the analytically continued
correlator this is not important unless the contour crosses a nonanalytic
region. Whatever singularities exist in the infrared region (Regions B,
B', or B'' in Fig.~\ref{figa}), the contour includes them. The resummation on
the contour is explicitly perturbative. For the resummation on the cut, the
extrapolation of the running of the coupling constant to the infrared region
is crucial since one has to interpret the integration over the infrared 
region. Formal
manipulations with $t=\beta_0\alpha(M_\tau^2)\ln(M_\tau^2/s)$ give
\begin{equation}\label{IrBorel}
\int_0^{M_\tau^2}\alpha(s)ds=\int_0^{M_\tau^2}
  \frac{\alpha(M_\tau^2)ds}{1+\beta_0\alpha(M_\tau^2)\ln(s/M_\tau^2)}
  =\frac{M_\tau^2}{\beta_0}\int_0^{\infty}
  \frac{e^{-t/\beta_0\alpha(M_\tau^2)}dt}{1-t}.
\end{equation}
Not all expressions in~(\ref{IrBorel}) are well-defined. This is particularly
true for the third form being
a Borel representation. The problem can be reformulated as a divergence of the
asymptotic series. Indeed, by expanding the expression for the running
coupling under the integration sign in a PT series one has
\begin{equation}\label{serIR}
\int_0^{M_\tau^2}\alpha(s)ds=\sum_nn!\pfrac{\beta_0\alpha(M_\tau^2)}\pi^n.
\end{equation}
The summation of the series in Eq.~(\ref{serIR}) is related to the
interpretation of the integral. Therefore, an integrable behaviour of the
coupling constant at small $s$ offers a recipe for the summation of the
asymptotic series. This solution is strongly model dependent because the
extrapolation of the evolution into the infrared region is essentially
arbitrary. The explicit form of the extrapolation in Eq.~(\ref{sigmadef})
gives an extrapolation motivated by analytic continuation. It can also be
considered as a special change of the renormalization
scheme~\cite{Krasnikov:1995is}. Indeed, for the coupling $a_M$ with the
evolution given in Eq.~(\ref{almev}) one obtains an infrared fixed point. 
Without the negative part of the spectrum that emerges in the exact treatment
of quantities originally defined in the Euclidean domain 
the analytic moments can be defined by
\begin{equation}\label{ancontmom}
m_{00}^{\rm anal}=\int_0^{M_\tau^2}\frac{\sigma_c(s)ds}{M_\tau^2}
\end{equation}
One can relate these
moments to the moments on the contour by the relation
\begin{equation}\label{ancontvsmymom}
m_{00}^{\rm anal}=-\frac1{\beta_0}e^{-\pi/\beta_0\alpha_\tau}+m_{00}
\end{equation}
which contains an explicit ``non-perturbative'' term.

The question of uniqueness of the resummation is important. However, it is only
relevant if many terms of the PT expansion are known. It happens that this 
uniqueness can be usefully employed in multi-loop
computations~\cite{Kazakov:1984km,Kazakov:1986mu}. Note that it is also
important to have examples of higher order PT behaviour in simple QFT models
as unique record-breaking computations
show~\cite{Kazakov:1979ik,Gorishnii:1983gp}. In some cases uniqueness is
good laboratory for checking multi-loop
techniques~\cite{Davydychev:1999ic,Kotikov:1990kg,Groote:1998ic}. The methods
are also applied in some exotic areas as nonlinear sigma model, topological
theories, and SUSY~\cite{Grisaru:1986wj}. 

\section{Summary}
We have reviewed different ways of interpreting perturbation theory results 
for the description of
$\tau$-decay observables. Experimental data are very precise and theory
matches it by unprecedented numbers of PT terms. Higher orders of PT are 
available for
many cases that allows both for the description of data and for the extraction
of the
parameters of the theory with high accuracy. However, the PT series converge
slowly requiring improvement that can be achieved through:
\begin{itemize}
\item[i)]manipulation with schemes for a set of related observables avoiding
  artificial intermediate quantities such as $\MSbar$ quantities;

\item[ii)]different resummation techniques;

\item[iii)]the avoidance of unnecessary expansions -- treating final
  polynomials in the coupling as exact expressions, e.g.\ $\beta$ function
  and anomalous dimensions in renormalization group equations for coefficient
  functions of operator product expansions

\end{itemize}
and some other improvements that you name. This richness is available due to 
RG properties that allows one to control the scaling behaviour and the
invariance of the theory.

\subsection*{Acknowledgement}
This work was supported by the grant of the Ministry of Education and Science
No.~8412, by the Estonian target financed project No.~0180056s09, and by the
Estonian Science Foundation under grant No.~8769. A.A.P.\ acknowledges the
partial support by RFFI grant 11-02-00112-a. S.G.\ acknowledges the support by
the Deutsche Forschungsgemeinschaft (DFG) under Grant No.~436~EST~17/1/06 and
by the Forschungszentrum of the Johannes-Gutenberg-Universit\"at Mainz
``Elementarkr\"afte und Mathematische Grundlagen (EMG)''.

\begin{appendix}
 
\section{Appendix}
\setcounter{equation}{0}\def\theequation{A\arabic{equation}}
The phenomenology of hadronic $\tau$ decays is contained in the correlator of
the weak currents
$j_{\mu}^W(x) = \cos(\theta_C) \bar{u}\gamma_{\mu}(1-\gamma_5) d
+\sin(\theta_C) \bar{u}\gamma_{\mu}(1-\gamma_5) s$,
\[
i\!\int\!\! \langle Tj_{\mu}^W(x)j_{\nu}^{W+}(0) \rangle e^{iqx}dx
=(q_\mu q_\nu - q^2 g_{\mu\nu})\Pi^{\rm had}(q^2)
\]
with $\rho(s)={\rm Im}~\Pi^{\rm had}(s+i0)/\pi$ and 
\[
\Pi^{\rm had}(q^2)=\int \frac{\rho(s)ds}{s-q^2}
\]

The total $\tau$-decay rate
\[
R_{\tau S=0}=\frac{\Gamma(\tau \rightarrow H_{S=0} \nu)}
{\Gamma(\tau \rightarrow l \bar{\nu} \nu )}
\sim \int_0^{M_\tau^2}\left(1-\frac{s}{M_\tau^2}\right)^2
\left(1+\frac{2s}{M_\tau^2}\right)\rho(s)ds
\]
is a useful observable which is measured with high experimental precision.
Since $\rho^{\rm had}(s)$ is a distribution one considers moments of the 
spectral density of the form
\[
M_{kl}=\frac{(k+l+1)!}{k!l!}\int_0^{M_\tau^2}
  \left(1-\frac{s}{M_\tau^2}\right)^k\pfrac{s}{M_\tau^2}^l
  \frac{\rho(s)ds}{M_\tau^2}\equiv 1+m_{kl}
\]
which contain information about the hadronic spectral density
$\rho^{\rm had}(s)$. The
function $\rho(s)$ is related to Adler's function by
\[
D(Q^2)=-Q^2\frac{d}{dQ^2}\Pi(Q^2)=Q^2\int\frac{\rho(s)ds}{(s+Q^2)^2} 
\]
where $Q^2=-q^2$ and $D(Q^2)$ are computable in PT. In the mass zero limit, the
PT expression for Adler's function reads
\[
D(Q^2)=1+a_s+k_1 a_s^2+k_2 a_s^3+k_3 a_s^4+O(a_s^5)
\]
with $a_s=\alpha_s(Q^2)/\pi$. The $\MSbar$-scheme coefficients read
\begin{equation}
k_1=\frac{299}{24}-9\zeta(3),\quad
k_2=\frac{58057}{288}-\frac{779}4\zeta(3)+\frac{75}2\zeta(5)
\end{equation}
while $k_3=49.08$~\cite{Gorishnii:1990vf,Baikov:2008jh}.
The correction to the total width $\delta_P$ is 
\begin{equation}\label{intro-full-width}
\delta_P^{\rm th} = a_s + 5.2023 a_s^2
  + 26.366 a_s^3 + (78.003 + 49.08) a_s^4 + O(a_s^5)
\end{equation}
which corresponds to the experimental value $\delta^{\rm exp}=0.216\pm0.005$.

The renormalization group equation for $a(\mu^2)$ reads
\begin{equation}
\mu^2\frac{da}{d\mu^2}=\beta(a)=-a^2(\beta_0+\beta_1a+\beta_2a^2
  +\beta_3a^3+\ldots\ )
\end{equation}
with
\begin{equation}
\beta_0=\frac94,\quad
\beta_1=4,\quad
\beta_2=\frac{3863}{384},\quad
\beta_3=\frac{140599}{4608}+\frac{445}{32}\zeta(3)
\end{equation}
for $N_c=n_f=3$~\cite{Tarasov:1980au,vanRitbergen:1997va}. 

The coefficients $f_{in}$ are given by
\begin{eqnarray}\label{eqn4}
f_{0n}&=&\tilde I(0,n),\quad
f_{1n}=\beta_0\tilde I(1,n),\quad
f_{2n}=\beta_0^2\Big(\tilde I(2,n)+\rho_1\tilde I(1,n)\Big),\nonumber\\
f_{3n}&=&\beta_0^3\Big(\tilde I(3,n)+(I_\tau(2)-I_\tau(1)^2-\frac13\pi^2)
  \tilde I(1,n)+\frac52\rho_1\tilde I(2,n)+\rho_2\tilde I(1,n)\Big),
  \nonumber\\
f_{4n}&=&\beta_0^4\Big(\tilde I(4,n)-3(I_\tau(2)-I_\tau(1)^2-\frac13\pi^2)
  \tilde I(2,n)\nonumber\\&&\qquad+2(I_\tau(3)-3I_\tau(1)I_\tau(2)
  +2I_\tau(1)^3)\tilde I(1,n)\\&&
  +\rho_1(\frac{13}3\tilde I(3,n)+5(I_\tau(2)-I_\tau(1)^2-\frac13\pi^2)
  \tilde I(1,n))+3\rho_2\tilde I(2,n)+\rho_3\tilde I(1,n)\Big)\nonumber
\end{eqnarray}
with
\begin{eqnarray}\label{eqn5}
I(m,n)&=&\frac{m!}{(n+1)^m},\quad
I_\tau(m)=2 I(m,0)-2 I(m,2)+I(m,3),\nonumber\\
I(m,n)&=&I(m)+\sum_{p=0}^m{m\choose p}I_\tau(p)\tilde I(m-p,n).
\end{eqnarray}

The $\rho_i$ are scheme-independent quantities given by
\begin{eqnarray}
\rho_1&=&\frac{\beta_1}{\beta_0^2},\quad
\rho_2=\frac{1}{\beta_0^3}\big[\beta_2-\beta_1k_1+\beta_0(k_2-k_1^2)\big],
\nonumber\\
\rho_3&=&\frac{1}{\beta_0^4}\big[\beta_3-2\beta_2k_1+\beta_1k_1^2
  +2\beta_0(k_3-3k_1k_2+2k_1^3)\big]
\end{eqnarray}

\end{appendix}
%%%%%%%%%%%%%%%End-of-App%%%%%%%%%

\end{document}